\documentstyle[preprint,aps]{revtex}

\begin{document}
\title{On Eigenvalues and Eigenfunctions Absent in the Actual Solid State Theory}
\author{P. Pereyra}
\address{F\'{i}sica Te\'{o}rica y Materia Condensada, UAM-Azcapotzalco, Av. S. Pablo\\
180, C.P. 02200, M\'{e}xico D. F., M\'{e}xico }
\date{\today}
\maketitle

\begin{abstract}
In this letter new, closed and compact analytic expressions for the
evaluation of resonant energies, resonant bound-states, eigenvalues and
eigenfunctions for both scattering and bounded $n$-cell systems are
reported. It is shown that for (scattering and bounded) 1-D systems the
eigenfunctions $\Psi _{\mu ,\nu }(z)$ are simple and well defined functions
of the Chebyshev polynomials of the second kind $U_{n}$, and the energy
eigenvalues $E_{\mu ,\nu }$ (in the $\mu $-th band) are determined by the
zeros of these polynomials. New insights on the energy gap and the
localization effect induced by phase coherence are shown.
\end{abstract}

\draft

Different techniques and approaches have been developed to determine, as
exactly as possible, eigenvalues and eigenfunctions. In quantum theory the
knowledge of these quantities represents a central aim and afford the
possibility of evaluating other quantities. It is well known that, except
for a small number of cases, it is not easy to obtain eigenvalues with the
precision that one would like\ to have and to deduce explicit analytical
expressions for the eigenfunctions. In the actual solid state physics
theory, the eigenfunctions and eigenvalues are in some sense beyond it, and
the theory is basically designed to describe systems in the continuous
spectrum limit. The famous and widely accepted Bloch function $%
e^{ik_{B}\cdot r}u_{\mu ,k}(r),$ strictly valid only for {\it infinite}
periodic systems[1], becomes the obvious and natural starting point for
almost any approach of periodic systems, even though the periodic function $%
u_{\mu ,k_{B}}(r)$ remains in general practically unknown. Calculations
based on the Bloch function lead also, almost automatically, to introduce
the reciprocal space that inexorably, aside from text books' didactic
presentations, may obscure the analysis. As a consequence, great efforts and
imagination had to be bestown to explain the vast world of mechanisms,
effects and properties discovered and populating the actual solid state and
condensed matter physics[2-6]. The allowed and forbidden energy bands, which
should not be confused with the energy eigenvalues of finite systems, are
usually determined after a complex and perhaps painful numerical
calculations with the aid of experimental input to fit parameters. On the
other hand, high precision experiments and energy dependent applications, as
those related to optical excitations, require one to solve the fine
structure in the bands. This has been so far difficult for the actual
theory. It is the purpose of this letter to extend the analytic methods of
the scattering theory, for finite periodic systems with arbitrary potential
shape, to determine compact and general expressions for an easy and simple
evaluation of the eigenvalues and eigenfunctions, which are neither the
energy bands nor the Bloch functions. In other words, the present approach
allows one to solve the fine structures in bands completely.

Concerning band structures and electronic properties, there is an abundant
literature and great contributions extending from the nearly free electron
model to pseudopotential methods and Local Density Approximations. Important
attempts were made forty years ago by Kohn to study the analytic properties
of Bloch waves and Wannier functions[2]. Since those years, the complexity
of involved Green's function and matching methods[7], together with a
complete zoo of effective masses, perturbative methods, envelope functions,
dispersion relations, and much more, have dominated the theory of real
periodic systems[8-11].

A completely different and natural approach to deal with finite (or locally)
periodic systems is being developed[12]. In this approach the Bloch
functions, reciprocal spaces, Brillouin zones, and other intricate concepts
and knotty theoretical structures {\it are not needed at all} in order to
calculate fundamental quantities. The theory of finite periodic systems
(TFPS), based on the transfer matrix method (originally proposed to
calculate scattering amplitudes, and related properties in superlattices),
will be expanded to include also the stationary properties and to deduce,
new and closed formulas for quantities as fundamental and basic as the
eigenvalues and eigenfunctions.

To encompass most of the various types of systems, I shall consider three
representative cases, distinguished by their boundary conditions. Hence, I
shall refer to open, bounded and quasi-bounded systems, as shown in figure
1. While for scattering systems (figure 1a)) I will refer to resonant
energies and resonant bound-states, for bounded systems I will talk of
energy eigenvalues and eigenfunctions. For the reasons of simplicity and
lack of space, I will restrict to the widely used one channel (1-D)
approximation. Some multichannel generalizations are almost
straightforwardly obtained[13].

It is well known, in the scattering theory, that resonant transmission
occurs precisely when the incident energy coincides with a bound-state
energy in the scatterer system, which, for our purpose, is locally periodic
with $n$ cells (of length $l_{c}$ each) in the transmission direction. For
this kind of systems, the transmission amplitudes of $N$-propagating modes ($%
N$-channels) are obtained from[12] 
\begin{equation}
t_{N,n}^{T}=\frac{1}{p_{N,n}-\beta ^{-1}\alpha \beta \ p_{N,n-1}}
\end{equation}
where the functions $p_{N,n}$ are $N\times N$ matrix polynomials fully
determined in terms of the single-cell transfer matrix $M$, which for time
reversal invariant systems has the structure 
\begin{equation}
M=\left( 
\begin{array}{cc}
\alpha & \beta \\ 
\beta ^{\ast } & \alpha ^{\ast }
\end{array}
\right) .
\end{equation}
In the one channel (one propagating mode) approximation $\alpha $ and $\beta 
$ are complex scalars and $p_{1,n}$ is the Chebyshev polynomial of the
second kind $U_{n}(\alpha _{R})$ evaluated at the real part of $\alpha .$
Because of the close relation between the resonant structure and the
spectral properties, we can determine simple expressions to evaluate the
resonant energies. If we use the identity $U_{n}U_{n-2}=U_{n-1}^{2}-1,$ the
whole $n$-cell system transmission coefficient can be rewritten as 
\begin{equation}
\left| t_{n}\right| ^{2}=\frac{\left| t\right| ^{2}}{\left| t\right|
^{2}+U_{n-1}^{2}(1-\left| t\right| ^{2})}.
\end{equation}
Here $\left| t\right| ^{2}$ is the single-cell transmission coefficient. It
is clear from this expression that the transmission resonances occur when
the polynomial $U_{n-1}$ becomes zero. Therefore, the $\nu $-th resonant
energy $E_{\mu ,\nu }$ is a solution of 
\begin{equation}
(\alpha _{R})_{\nu }=\cos \frac{\nu \pi }{n}
\end{equation}
where $(\alpha _{R})_{\nu }$ is the $\nu $-th zero of the Chebyshev
polynomial and $\nu =1,2,...n-1$. The index $\mu $ labels the bands,
peculiar to periodic systems and entirely determined by the phase coherence.
In the transfer matrix approach the allowed energy bands are those energies
which satisfy the condition $\left| \alpha _{R}\right| \leq 1$. To
illustrate the use of equation (4), let us consider the square-barrier
superlattice of figure 1a). This is similar to the familiar Kronig-Penney
model (except that here we have a finite number of cells $n$) with barrier
height $V_{o}$, and valley and barrier widths $a$ and $b$, respectively. In
this case 
\begin{equation}
\cos k_{\nu }a\cosh q_{\nu }b-\frac{k_{\nu }^{2}-q_{\nu }^{2}}{2k_{\nu
}q_{\nu }}\sin k_{\nu }a\sinh q_{\nu }b=\cos \frac{\nu \pi }{n}
\end{equation}
\ with $k_{\nu }^{2}=2m_{v}^{\ast }E_{\mu ,\nu }/\hbar ^{2}$ and $q_{\nu
}^{2}=2m_{b}^{\ast }(V_{o}-E_{\mu ,\nu })/\hbar ^{2}$. Notice that each
energy band contains the same number of resonant energies as confining wells
have the scatterer periodic system. In this case is $n-1$. In figure 2, some
of these energies and the associated level densities $\rho (E)$ are plotted
for different values of $n$. The level density behavior tends rapidly, as a
function of $n,$ to that of the Kronig-Penney model, though the continuous
spectrum limit is reached only when $n$ $\rightarrow \infty $.

An important extension of the scattering approach, and the transfer\ matrix
method, is to studying stationary properties of bounded periodic systems. If
we have systems like those shown in figures 1b-c), we can apply the main
results of the TFPS and introduce the boundary conditions. For a periodic $n$%
-cell system (with length $nl_{c}$) bounded by infinite hard walls, it is
easy to show that the energy eigenvalues are determined from 
\begin{equation}
(\alpha _{n}-\alpha _{n}^{\ast }+\beta _{n}^{\ast }-\beta _{n})=0
\end{equation}
which, using the relations[12] $\alpha _{n}=U_{n}-\alpha ^{\ast }U_{n-1}$
and $\beta _{n}=\beta U_{n-1}$can be written as $U_{n-1}(\alpha _{I}-\beta
_{I})=0$. Here the subscript $I$ refers to the imaginary part. Notice that $%
n-1$ of the energy eigenvalues of the bounded system are the zeros of the
Chebyshev polynomial $U_{n-1}$. For the particular example shown in figure
1b), which length is $nl_{c}+a,$ the eigenvalues are obtained from $%
U_{n}\sin ka+(\alpha _{I}\cos ka-\alpha _{R}\sin ka-\beta _{I})U_{n-1}=0$.
These eigenvalues are used below when evaluating the eigenfunctions.

Another type of system that I would also like to consider is a periodic $n$%
-cell system confined by finite potential walls as shown in figure 1c). In
this case, assuming $E<V_{w}$, the eigenvalues are obtained from 
\begin{equation}
h_{w}U_{n}-f_{w}U_{n-1}=0
\end{equation}
with $f_{w}=(\alpha _{R}-\alpha _{I}\frac{q_{w}^{2}-k_{o}}{2q_{w}k_{o}}%
+\beta _{I}\frac{q_{w}^{2}+k_{o}}{2q_{w}k_{o}})$ and $h_{w}=(1+\frac{%
q_{w}^{2}-k_{o}}{2q_{w}k_{o}})$. Here $q_{w}^{2}=2m(V_{w}-E)/\hbar ^{2}$ and 
$k_{o}$ is the wave vector at $z=z_{o}+0^{+}$. Below, these eigenvalues are
used to evaluate the corresponding eigenfunction for an specific system.

I shall now refer to the eigenfunctions $\Psi _{\mu ,\nu }(z)$ of finite
periodic systems. For this purpose, it is convenient first to fix some
notation. As shown in figure 1, the coordinates $\left\{ z_{j}\right\} ,$
with $j=0,1,2,...n$, define a set of points separated by multiples of $l_{c}$%
, i.e. $z_{j}=jl_{c}.$ If we are interested in evaluating functions at any
point $z$ in the $j$-th cell, it is useful to define the difference $\delta
z=z-z_{j}\leq l_{c}$. Taking into account the transfer matrix multiplicative
properties, the total transfer matrix $M_{T}(z_{o}\rightarrow z)$ relating
the wave vectors $\Phi (z_{o})$ and $\Phi (z)$, can be factorized as $%
M_{j}(z_{o}^{\prime }\rightarrow z)M_{o\rightarrow o^{\prime
}}(z_{o}\rightarrow z_{o}^{\prime })$, with $z_{o}^{\prime }=z_{o}+\delta z$%
, and subsequently, the wave vector can be written as 
\begin{equation}
\Phi (z)=\left( 
\begin{array}{c}
\overrightarrow{\varphi }(z) \\ 
\overleftarrow{\varphi }(z)
\end{array}
\right) =\left( 
\begin{array}{cc}
\alpha _{j} & \beta _{j} \\ 
\beta _{j}^{\ast } & \alpha _{j}^{\ast }
\end{array}
\right) \left( 
\begin{array}{c}
\overrightarrow{\varphi }(z_{o}^{\prime }) \\ 
\overleftarrow{\varphi }(z_{o}^{\prime })
\end{array}
\right) .
\end{equation}
Here $\alpha _{j}=U_{j}-\alpha ^{\ast }U_{j-1}$, $\beta _{j}=\beta U_{j-1}$,
and the functions $\overrightarrow{\varphi }(z)$ and $\overleftarrow{\varphi 
}(z)$ represent, depending on the difference $E-V(z)$, the right and left
propagating or exponentially increasing and decreasing functions,
respectively[12]. The eigenfunctions are easily obtained by using these
relations, the boundary conditions and the general expressions for $\alpha
_{j}$ and $\beta _{j}$.

For a scattering system like the one shown in Fig. 1a), the wave function at 
$z$\ is given by 
\begin{equation}
\Psi (z,E)=\overrightarrow{\varphi }(z_{o}^{\prime })\left[ \alpha
_{j}+\beta _{j}^{\ast }-(\alpha _{j}^{\ast }+\beta _{j})\frac{\beta
_{n}^{\ast }}{\alpha _{n}^{\ast }}\right]
\end{equation}
It is clear that evaluating the function $\Psi (z,E)$ at $E_{\mu ,\nu }$ we
have the corresponding $\nu $-th resonant bound state in the $\mu $-th band 
\begin{equation}
\Psi _{\mu ,\nu }(z)=\Psi (z,E_{\mu ,\nu }).
\end{equation}
In figures 3a-d) we plot four different functions for a system with the
potential parameters of the superlattice $GaAs(Al_{0.3}Ga_{0.7}As/GaAs)^{12}$%
. In 3a) we have the function $\left| \Psi _{2}(z,E)\right| ^{2}$ for an
arbitrary energy $E$ ($\neq E_{\mu ,\nu }$) within the second energy band.
This is an extended aperiodic wave function with a complicated behavior
along the superlattice. In figures 3b and 3c), we have instead the functions 
$\left| \Psi _{2,2}(z)\right| ^{2}$ and $\left| \Psi _{4,3}\right| ^{2}$%
corresponding to the second and third resonant energies in the second ($\mu
=2$) and fourth ($\mu =4$) energy bands. In these cases the bound-state
functions are modulated by an oscillating envelope function with $\nu -1$\
minima plus those at $z_{o}=0$ and $z_{n}=nl_{c}$. Notice that, since these
wave functions describe not only extended but also transmitted states, they
are, except for a few points in $\left| \Psi _{2,2}(z)\right| ^{2},$
different from zero with the probability to find the particle at the two
ends of the system different from zero. This will change for bounded
systems, of course. Finally, in figure 3d) we plot $\left| \Psi (z,E)\right|
^{2}$ at some point in the gap between the second and third bands. The wave
function behavior, for an incident energy in the gap, is not only compatible
with the well known vanishing of the transmission coefficient, it makes also
evident the localization effect induced by the phase coherence at the energy
gap, as suggested in some sense by Kohn[2]. This is a very appealing result
which may deserve further reflections.

To complete this discussion we shall\ now briefly refer to the other two
types of systems. Consider first the system bounded by infinite hard walls.
In this case, the $\mu ,\nu $ eigenfunctions are evaluated from 
\begin{equation}
\Psi _{\mu ,\nu }(z)=\overrightarrow{\varphi }(z_{0}^{\prime })(\alpha
_{j}+\beta _{j}^{\ast }-\alpha _{j}^{\ast }-\beta _{j})
\end{equation}
with $E=E_{\mu ,\nu }$ and the matrix elements $\alpha _{j}$ and $\beta _{j}$
as mentioned above are simple functions of the Chebyshev polynomials of
order $j$ and $j-1$. Care has to be taken when extra pieces of length $a/2$
are added to the system of length $nl_{c}$. For the particular case shown in
figure 1b) the last two terms in Eq. (11) have to be multiplied by $(\alpha
_{n}+\beta _{n}^{\ast }e^{-ika})/(\beta _{n}+\alpha _{n}^{\ast }e^{-ika}).$
Two of these eigenfunctions ($\left| \Psi _{2,2}(z)\right| ^{2},$ $\left|
\Psi _{2,3}(z)\right| ^{2}$) are plotted in figures 4a-b).Contrary to the
scattering system, the wave function vanishes at the ends. An important
difference between a periodic potential flanked by infinite hard walls and a
constant potential flanked also by infinite hard walls, is the presence (in
the periodic potential case) or absence (in the single well case) of phase
interference effect responsible for energy splittings that in the former
case develop the band structure and the energy eigenvalues.

Finally, for the system with finite walls mentioned above, the
eigenfunctions are given by 
\begin{equation}
\Psi _{\mu ,\nu }(z)=\overrightarrow{\varphi }(z_{0}^{\prime })(\alpha
_{jR}+\beta _{jR}+\frac{q_{w}}{k}(\beta _{jI}-\alpha _{jI}))
\end{equation}
Here$E_{\mu ,\nu }$ is obtained from equation (7). Some of these
eigenfunctions are plotted in figures 5a-c). Although imperceptible, the
wave functions decrease exponentially inside the potential walls. As in
previous figures, we distinguish two main characteristics: a strongly
oscillating behavior, which frequency grows with $\mu $ and $\nu $, and a
periodic modulation, with period $nl_{c}/\nu $.

In this letter various new and general formulas for the evaluation of
fundamental quantities like the resonant energies, bound states, energy
eigenvalues and eigenfunctions of finite periodic systems, have been
reported. Specific and illustrative examples and results, for both
scattering and bounded one-dimensional systems, have been also presented.

I acknowledge Project 29026-E from CONACyT, Mexico and the ICTP, Trieste
Italy. I benefited from comments and discussions with Professors. H.
Simanjuntak and N. March.

\begin{figure}[tbp]
\caption{Potential profiles, along the direction $z$, for open, bounded and
quasi-bounded $n$-cell systems.}
\end{figure}
\begin{figure}[tbp]
\caption{Level density in the first band of finite and infinite
(Kronig-Penney) $GaAs\left(Al_{0.3}Ga_{0.7}As/GaAs\right)^n$ superlattice
with $a=100 nm$, $b=30 nm$ and $V_o=0.23 eV$. Some eigenvalues $E_{1,\protect%
\nu}$ (in $eV$) for $n=8$ are also indicated.}
\end{figure}
\begin{figure}[tbp]
\caption{Squared wave functions for a system like in Fig. $1a)$ but with
square barrier potential and parameters as in Fig. $2$. In $a)$ extended $%
\left| \Psi_{\protect\mu}(z,E)\right| ^{2}$ with $E \neq E_{\protect\mu,%
\protect\nu}$ and $\protect\mu=2$ is shown; In $b)$, and $c) $ the indicated
resonant states; In $d)$ a localized wave function in the second gap is
shown.}
\end{figure}

\begin{figure}[tbp]
\caption{Eigenfunctions $\left| \Psi_{\protect\mu,\protect\nu}(z)\right|
^{2} $ for a system like in Fig. $1b)$ and parameters as in Fig. $2$. The
oscillations frequency and enveloping minima depend on the band and
excitation energy indices $\protect\mu$, $\protect\nu$. All functions vanish
at the surface. In $a)$ and $b)$ the particle density beneath the surface is
relatively high. }
\end{figure}
\begin{figure}[tbp]
\caption{Eigenfunctions in the lower bands for a system like in Figs. 1c)
but with square barriers and parameters as in Fig. 2), flanked by finite
walls.}
\end{figure}

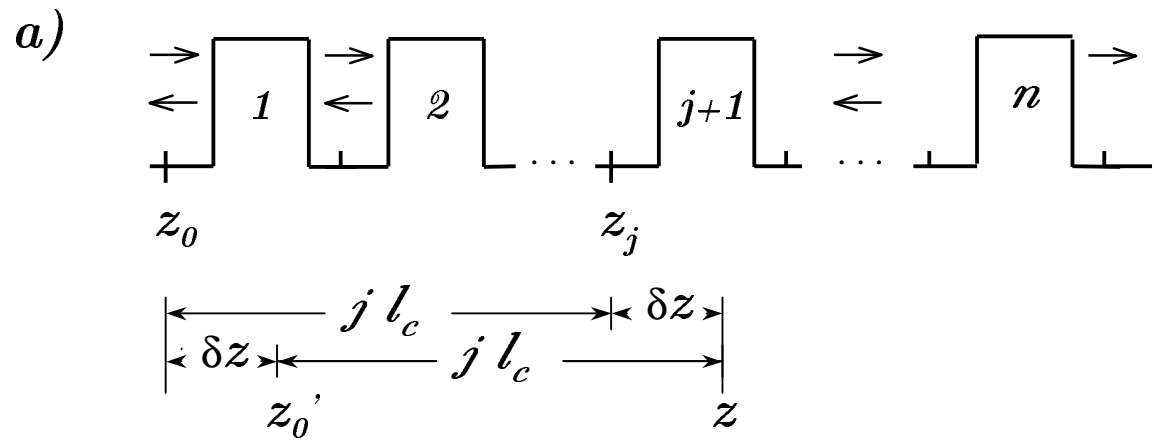
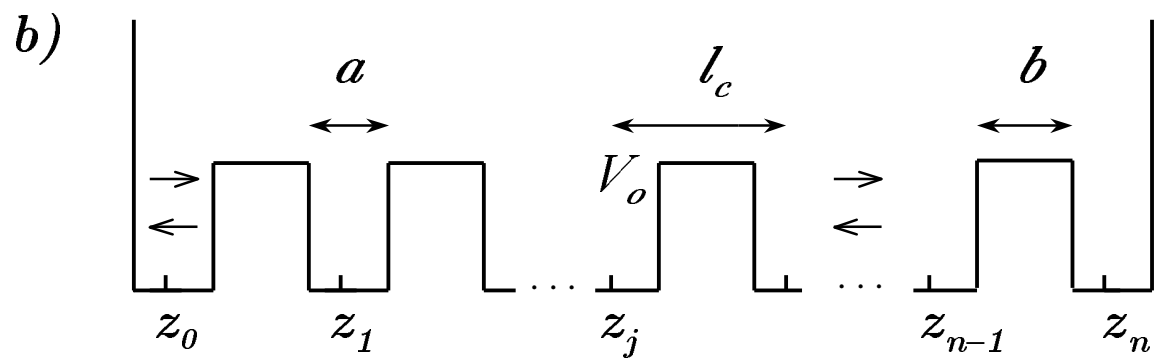
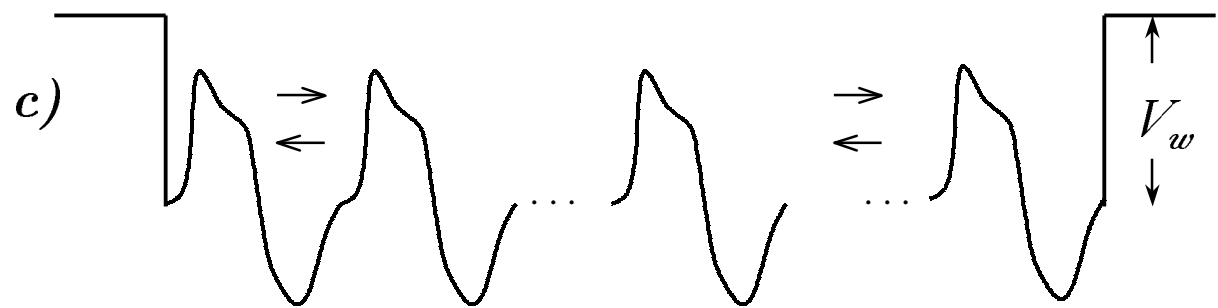
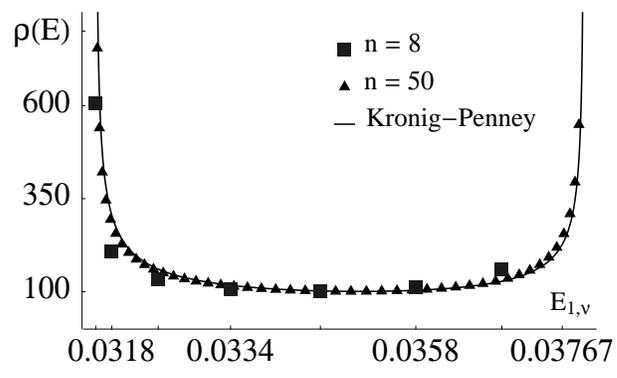
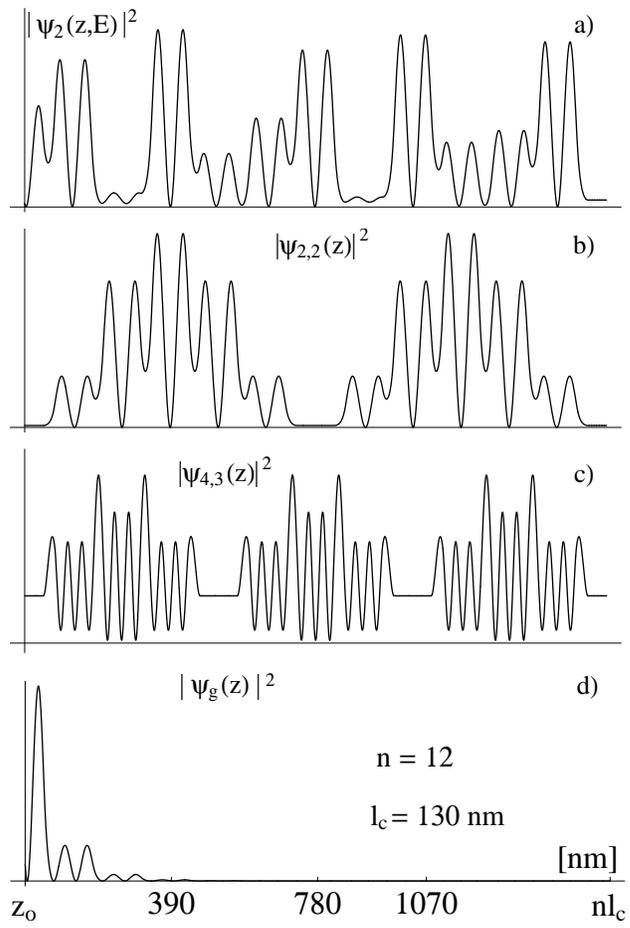
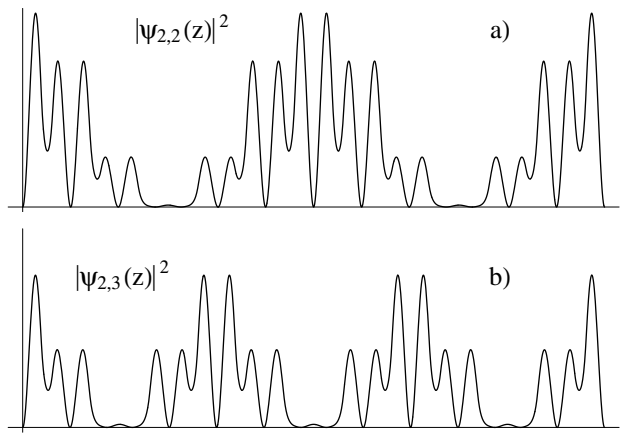
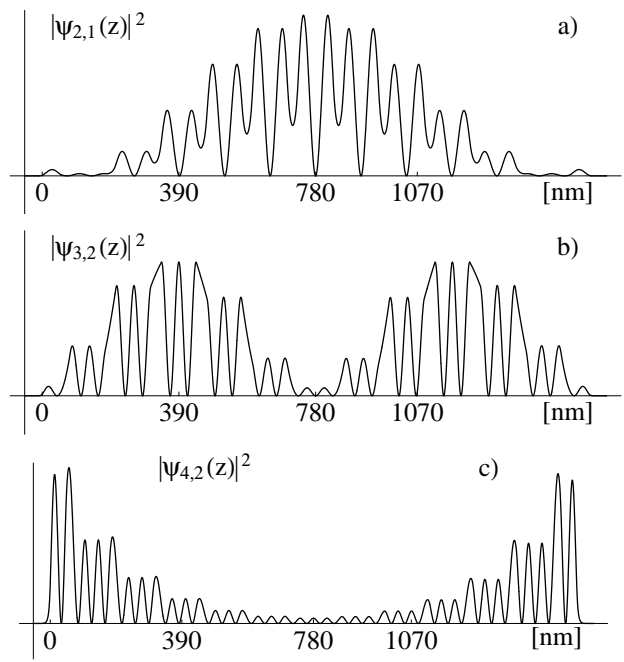
\end{document}